\documentclass[twocolumn]{aastex61}
\usepackage{graphicx}
\usepackage{amssymb}
\usepackage{amsmath,wasysym}
\usepackage{natbib}

\shorttitle{Proxima Centauri dynamo}
\shortauthors{Yadav et al.}

\begin{document}

\title{Magnetic cycles in a dynamo simulation of fully convective M-star Proxima Centauri}

\correspondingauthor{Rakesh K. Yadav}
\email{rakesh.yadav@cfa.harvard.edu}

\author{Rakesh K. Yadav}
\affil{Harvard-Smithsonian Center for Astrophysics, 60 Garden Street, 02138 Cambridge, USA}

\author{Ulrich R. Christensen}
\affil{Max-Planck-Institut f\"{u}r Sonnensystemforschung, Justus-von-Liebig-Weg 3, 37077 G\"{o}ttingen, Germany}

\author{Scott J. Wolk}
\affil{Harvard-Smithsonian Center for Astrophysics, 60 Garden Street, 02138 Cambridge, USA}

\author{Katja Poppenhaeger}
\affil{Astrophysics Research Center, Queen's University Belfast, BT7 1NN Belfast, United Kingdom}
\affil{Harvard-Smithsonian Center for Astrophysics, 60 Garden Street, 02138 Cambridge, USA}

\begin{abstract}

The recent discovery of an Earth-like exoplanet around Proxima Centauri has shined a spot light on slowly rotating fully convective M-stars. When such stars rotate rapidly (period $\lesssim 20$ days), they are known to generate very high levels of activity that is powered by a magnetic field much stronger than the solar magnetic field. Recent theoretical efforts are beginning to understand the dynamo process that generates such strong magnetic fields. However, the observational and theoretical landscape remains relatively uncharted for fully convective M-stars that rotate slowly. Here we present an anelastic dynamo simulation designed to mimic some of the physical characteristics of Proxima Centauri, a representative case for slowly rotating fully convective M-stars. The rotating convection spontaneously generates differential rotation in the convection zone which drives coherent magnetic cycles where the axisymmetric magnetic field repeatedly changes polarity at all latitudes as time progress. The typical length of the `activity' cycle in the simulation is about nine years, in good agreement with the recently proposed activity cycle length of about seven years for Proxima Centauri. Comparing our results with earlier work, we hypothesis that the dynamo mechanism undergoes a fundamental change in nature as fully convective stars spin down with age.

\end{abstract}


\section{Introduction} \label{sec:intro}
Observations have revealed a tight correlation between the stellar rotation period and the stellar activity. It is clear that rapidly rotating cool stars are more active than their slowly rotating counterparts. The relative magnitudes of the various magnetic activity indicators, e.g.~H$\alpha$ emission \citep[e.g.][]{reiners2012b, newton2016b}, Ca $\mathrm{II}$ H\&K emission \citep[e.g.][]{astudillo2016}, UV emission \citep[e.g.][]{stelzer2016}, and X-ray emission \citep[e.g.][]{jeffries2011, wright2011}, follow a robust trend as a function of the {\em Rossby} number $Ro=P/\tau$ ($P$ is the stellar rotation period and $\tau$ is the typical time scale of stellar convection). For $Ro\lesssim 0.1$, the activity indicators are typically saturated to a plateau, while, for $Ro>0.1$, their magnitude gradually decreases. This is commonly referred to as the `rotation-activity' relationship. Stellar activity is powered by the stellar magnetic field that also follows a qualitatively similar trend \citep{reiners2009, vidotto2014}. Magnetic fields in stars are generated by a dynamo mechanism working in their convection zone. Therefore, the rotation-activity relationship indirectly describes how the stellar dynamo behaves with stellar rotation.

In the context of stellar activity M-stars have a justifiably special place. They are the most numerous stars in our galaxy. Among cool stars, M-stars are the most active and many of them are known as `flare stars' for producing frequent flares \citep[e.g., see][]{west2004,  vida2016}. Furthermore, M-stars are particularly important for finding habitable exoplanets due to their smaller size that provides a better signal to noise ratio for detecting Earth-like exoplanets \cite[e.g.][]{irwin2008}. Since exoplanets with liquid water will orbit much closer than 1AU around M-stars, the activity levels in M-stars will have a crucial, if not governing, influence on the habitability of such exoplanets \citep[e.g.][]{cohen2014}. 

M-stars with mass less than about 35\% of the solar mass are believed to be fully convective, i.e.~these stars do not have a radiative core and a tachocline like region. Although the data are relatively scarce for FC M-stars with slow rotation periods, many studies have suggested that FC M-stars also follow a rotation-activity relationship similar to the stars with a radiative core, i.e.~a saturated activity below a threshold $Ro$ and a gradual decline for higher $Ro$ \citep{kiraga2007,reiners2009, jeffries2011,newton2016b,  astudillo2016, stelzer2016,wright2016}. Therefore, one may postulate that the rotation-activity relationship might be immune to the internal structure of stars. If confirmed with later observation, then this conjecture will be a stringent constraint on the basic stellar dynamo theory. Along with carrying out detailed observations of FC M-stars in the slowly rotating regime, it is imperative that we develop theoretical models that make sense of current and upcoming observations. The immediate need for this exercise is underscored by the discovery of a possible Earth-like exoplanet \citep{anglada2016} around Proxima Cantauri (Prox Cen). This is a frequently flaring, slowly rotating, FC M5.5 star with a rotation period of about 83 days \citep{kiraga2007, mascareno2016}. There are now several observational studies that claim activity cycles on Prox Cen. An earlier reporting \citep{cincunegui2007} suggested an activity cycle length of about 1.2 yr. Recent work based on a longer data set could not confirm the 1.2 yr period but found strong evidence for cycles with a length of approximately 7 years \citep{mascareno2016, wargelin2016}.

To understand the dynamo mechanism in fast rotating FC M-stars we recently performed a fully-nonlinear turbulent dynamo simulation \citep{yadav2015apjl}. The rotation rate was about 20 days which is fast enough to push FC M-stars to the saturated regime of the rotation-activity relationship \citep{reiners2014, wright2016}. The simulation self-consistently produces very strong magnetic fields, reaching several kilo Gauss, on both large and small length scales. The morphological features of the magnetic field also resemble the observations to a good extent \citep{yadav2015apjl}. Motivated by these favorable results we advance this simulation further using a slower rotation rate. As we show below, the increased Rossby number due to slower rotation rate leads to a fundamental change in the dynamo solution, and, instead of a quasi-steady dipole-dominated field, the simulation produces regular magnetic cycles.

\section{Simulation setup}
The simulation we present here is a slowly rotating version of the simulation we reported in \citet{yadav2015apjl}. For the sake of completeness, we repeat some of the relevant model details; further information can be found in \citet{yadav2015apjl} and \cite{yadav2015spots}. We employ the magnetohydrodynamic equations modified under the anelastic approximation \citep{lantz1999}. This approach is now widely used for simulating subsonic convection in the interiors of stars and planets \cite[e.g., see][]{browning2008, gastine2013zonal, augustson2015, duarte2016}. To model the convection zone of a star we consider a spherical shell that is bounded by inner radius $r_i$ and outer radius $r_o$. The depth of the convection zone is $D=r_o-r_i$. Due to technical limitation we cannot model a fully convective star with $r_i=0$. Therefore, we exclude a small region around the stellar center such that $r_i/r_o$=0.1. The convection is driven by an entropy contrast $\Delta s$ between the two boundaries. The simulation domain incorporates five density scale heights. The gravity varies linearly with radius. Both boundaries are stress-free for velocity, have constant entropy, and are insulating for the magnetic field. 

We use the open source\footnote{{\tt https://github.com/magic-sph}} code \texttt{MagIC} for this simulation. This code has been rigorously tested with community benchmarks \citep{jones2011, gastine2012b}. It uses spherical harmonic decomposition in the latitude and longitude direction and Chebyshev polynomial decomposition in the radial direction. The code also uses \texttt{SHTns}, an open source library for performing fast spherical harmonic transforms \citep{schaeffer2013}. We perform most of the temporal evolution of the simulation on a grid with 1024 points in longitude, 512 points in latitude, and 121 points in radius. 

\texttt{MagIC} solves the equations in non-dimensional form. The fundamental parameters that govern the system and the values chosen for them are the Prandtl number $P_r=\nu/\kappa$=0.1, the magnetic Prandtl number $P_m=\nu/\lambda$=0.2, the Ekman number $E=\nu D^{-2}\Omega^{-1}$ = $10^{-5}$, and the Rayleigh number $Ra=g_o\,D^3\,\Delta s (c_p\nu\kappa)^{-1}$ = $1.5\times10^9$, where $\nu$ is viscosity, $\kappa$ is thermal diffusivity, $\lambda$ is magnetic diffusivity, $\Omega$ is rotation frequency, $g_o$ is gravity at $r_o$, and $c_p$ is specific heat at constant pressure. Note that due to the higher $Ra$ value we use here, the magnetic Reynolds numbers ($u\,D/\lambda$, $u$ is local velocity) would be substantially higher in this simulation as compared to those in \cite{yadav2015apjl} if we keep the same $P_m$.  Therefore, in order to probe similar magnetic Reynolds numbers we decreased $P_m$ by a factor of 10 in this study.

The results of a simulation, obtained in non-dimensional terms, can be scaled differently to actual stars. Here we pick Prox Cen as an example of a slowly rotating fully convective star and equate the shell thickness $D$ of our model with 0.85 of the observed radius (0.141$ R_{sun}$). For viscosity and the diffusivities we must use much larger than realistic values in order to match the numbers of our non-dimensional parameters. In addition, we can fix either the rotation rate or the luminosity to the actual value, but not both simultaneously. Because here we are mainly interested in the correct scaling of time to physical units, we follow the approach by \cite{dobler2006}, by fixing $\Omega$ to the stellar value. This leaves us in the model with 0.3 times the actual luminosity of Prox Cen (which is 0.0017 $L_{sun}$). The outer radius of the model at $r_o$ = 93,700 km is 5\% below the stellar photosphere. With this choice the density contrast of 150 across our model shell agrees with that obtained in a simple polytropic star model with the same mass and radius as Prox Cen. Density and gravity at the outer model boundary are 2440 $kg/m^3$ and 1850 $m/s^2$. For the various diffusion coefficients, taken to be constant with radius, we have $\nu$=62,000 $m^2/s$, $\kappa$=620,000 $m^2/s$ and $\lambda$=310,000 $m^2/s$.

\section{Results}
\begin{figure}
\epsscale{1.2}
\plotone{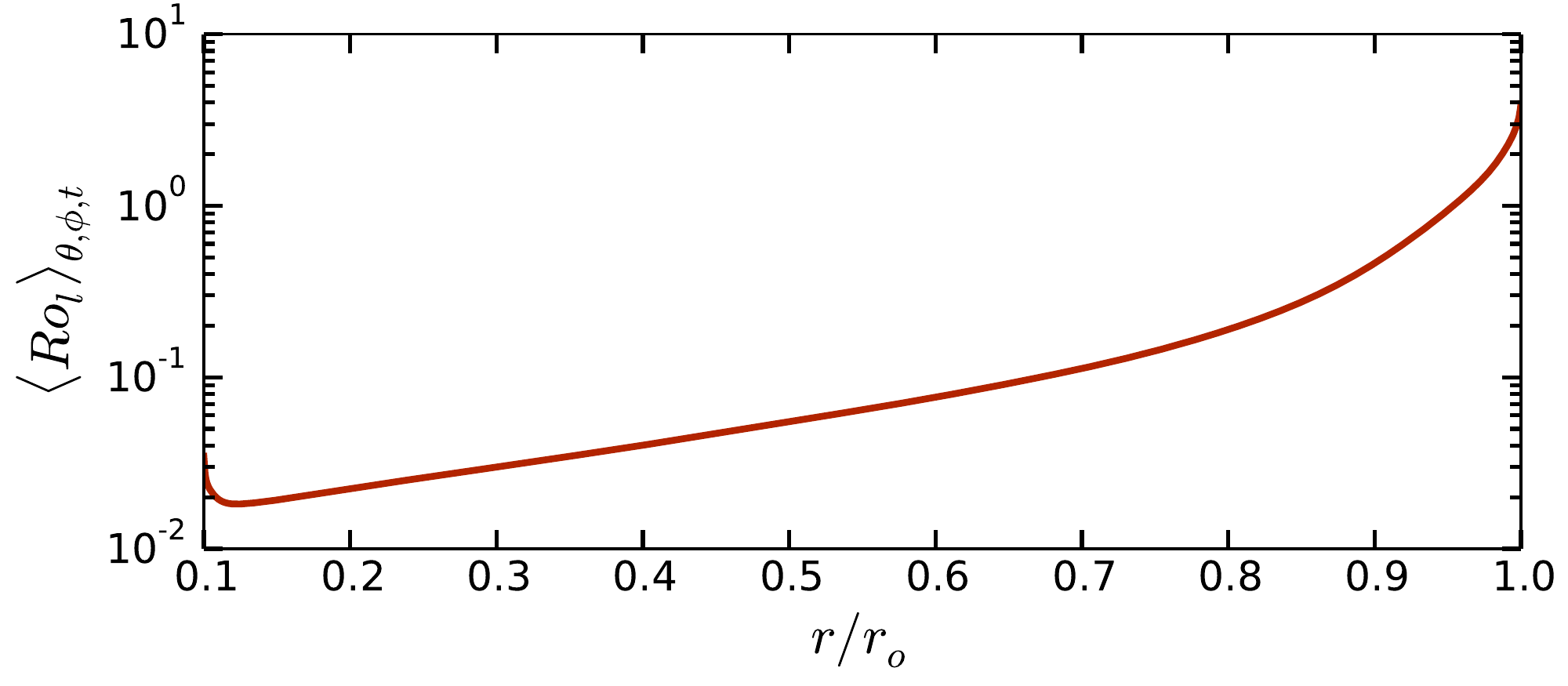}
\caption{ Radial variation of the local Rossby number averaged in longitude, latitude, and in time (about 100 rotations). \label{figRol}}
\end{figure}

To quantify the effect of rotation on convection we use the {\em local} Rossby number $Ro_l=u\Omega/l$, where $u$ is local velocity (in the rotating frame of reference) and $l$ is the local convection length scale \citep[for more details, see][]{christensen2006, yadav2015spots}. The time-averaged mean $Ro_l$ in our earlier faster rotating simulation is about 0.05. Due to the slower rotation rate in the simulation  we present here, the $Ro_l$ is larger with a value of about 0.25. Furthermore, as typically reported in density stratified convection studies \citep[e.g.][]{browning2008}, the value of $Ro_l$ varies in radius as shown in Fig.~\ref{figRol}.

\begin{figure*}
\epsscale{1}
\plotone{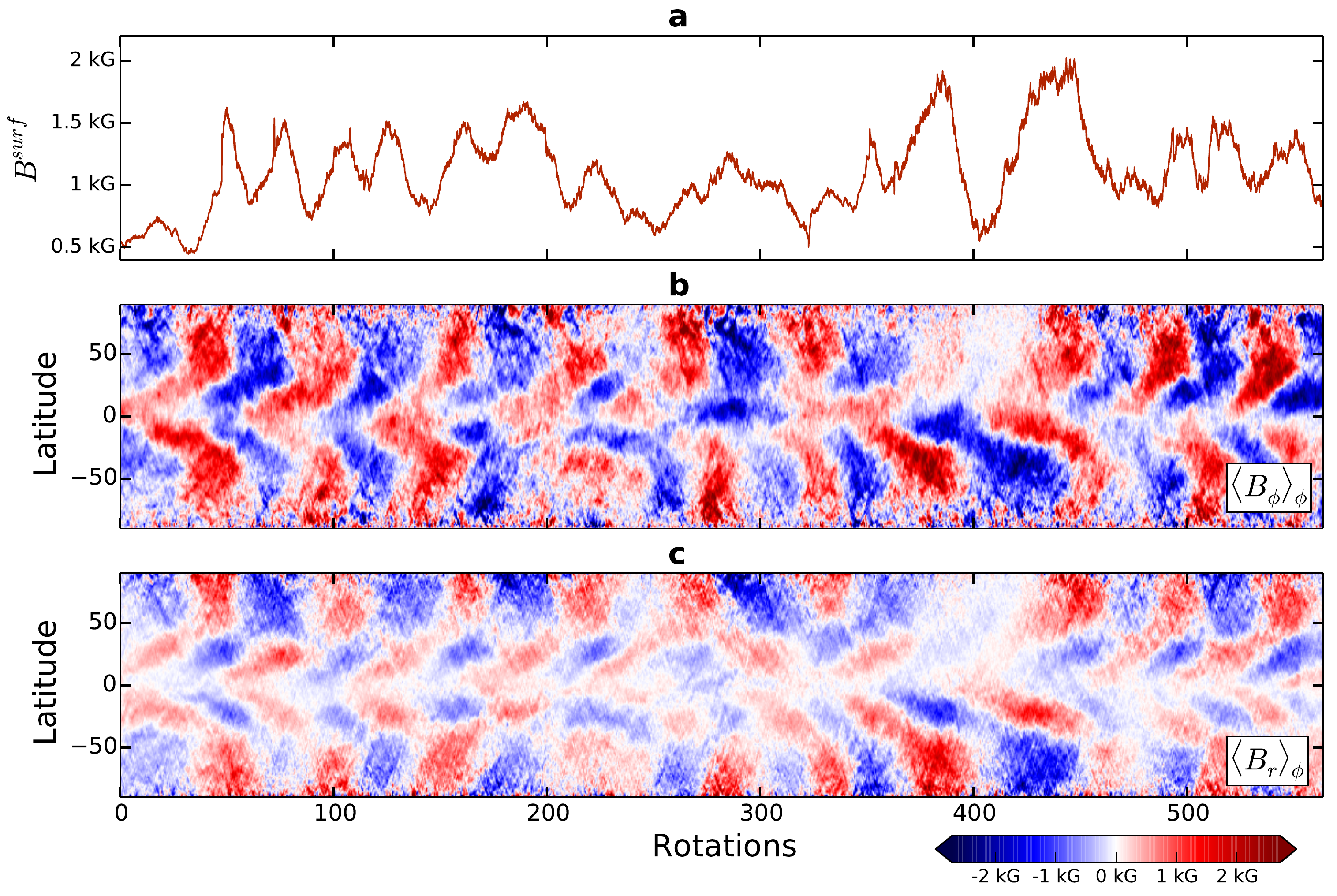}
\caption{Panel (a): Area averaged mean magnetic field strength on the simulation surface. Panel (b): Temporal evolution of the longitudinally averaged longitudinal magnetic field at a deeper radial level ($r=0.9r_o$). Panel (c): Longitudinally averaged radial magnetic field at a deeper radial level ($r=0.9r_o$). \label{fig1}}
\end{figure*}

The long term evolution of the simulation, capturing about 550 rotations\footnote{Equivalent to about 0.2$\tau_{mag}$, where $\tau_{mag}=D^2/\lambda$ is the magnetic diffusion time.}, is plotted in Figure \ref{fig1}(a). The panel displays the unsigned mean magnetic field $B^{surf}$ on the simulation surface. The time-averaged $B^{surf}$ is about 1.1 kG, although it varies\footnote{We again note that the dimensionalization procedure is not unique. If, instead of assuming the correct rotation rate we use the correct luminosity of Prox Cen, then the dimensional mean field strength is about 1.7 kG (using the scaling law by \cite{christensen2009}). The resulting rotation period would be about 53 days.} from about 0.5 kG to 2 kG. These field values are comparable to the observed (using Zeeman broadening) range of field strength in Prox Cen which is from 450 to 750 Gauss \citep{reiners2008}. We remind the reader that the outer surface of our simulation is a layer 5\% below the photosphere. Therefore, $B^{surf}$ in our simulation is likely larger than the photospheric value. There are distinct and sustained modulations present in the mean magnetic field in Figure \ref{fig1}(a). Applying the Lomb-Scargle periodogram to the entire time series gives a peak at around 40 rotations, meaning that the magnetic field strength peaks on-average every 9 years.

\begin{figure*}
\epsscale{0.15}
\plotone{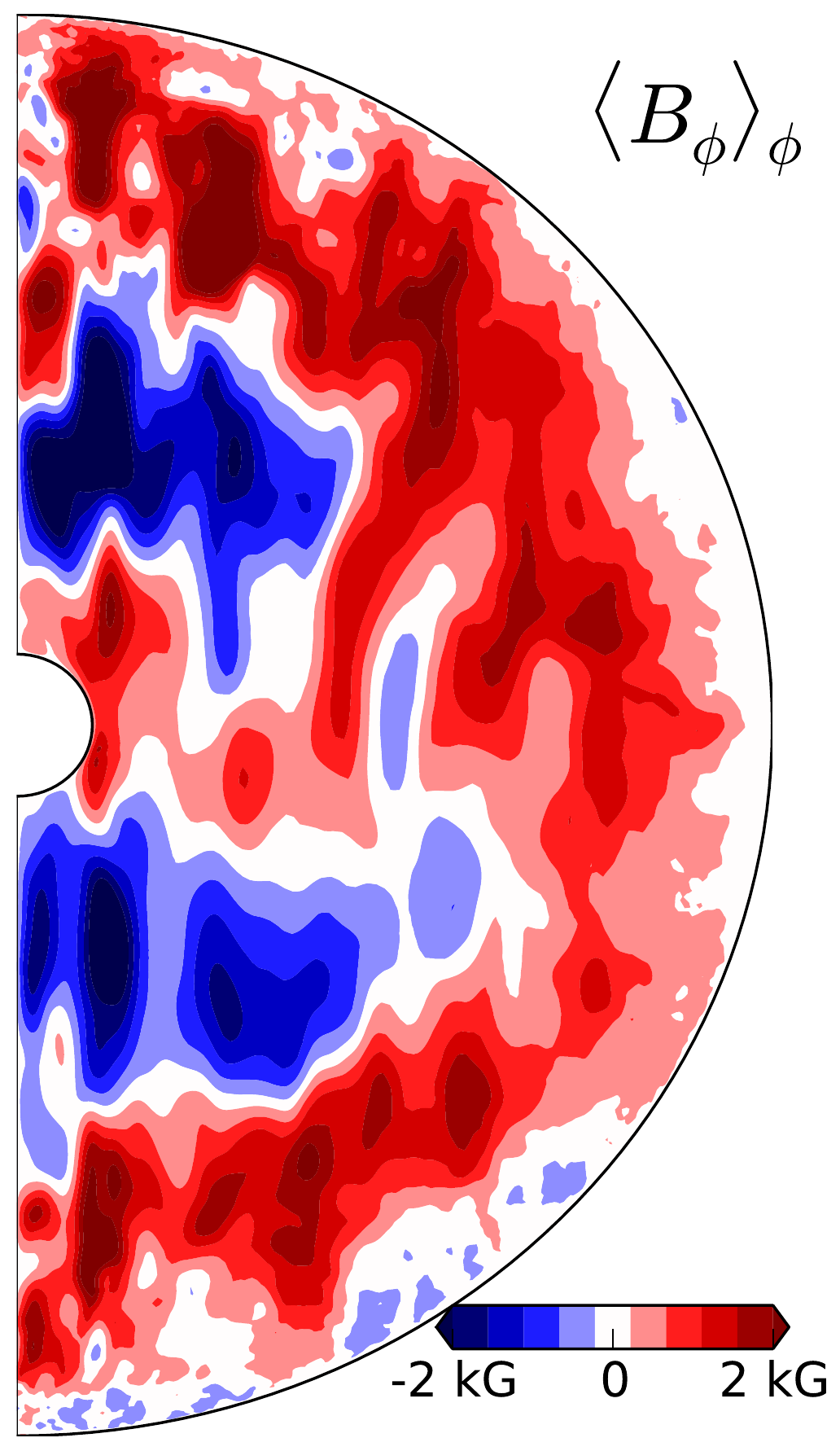} \plotone{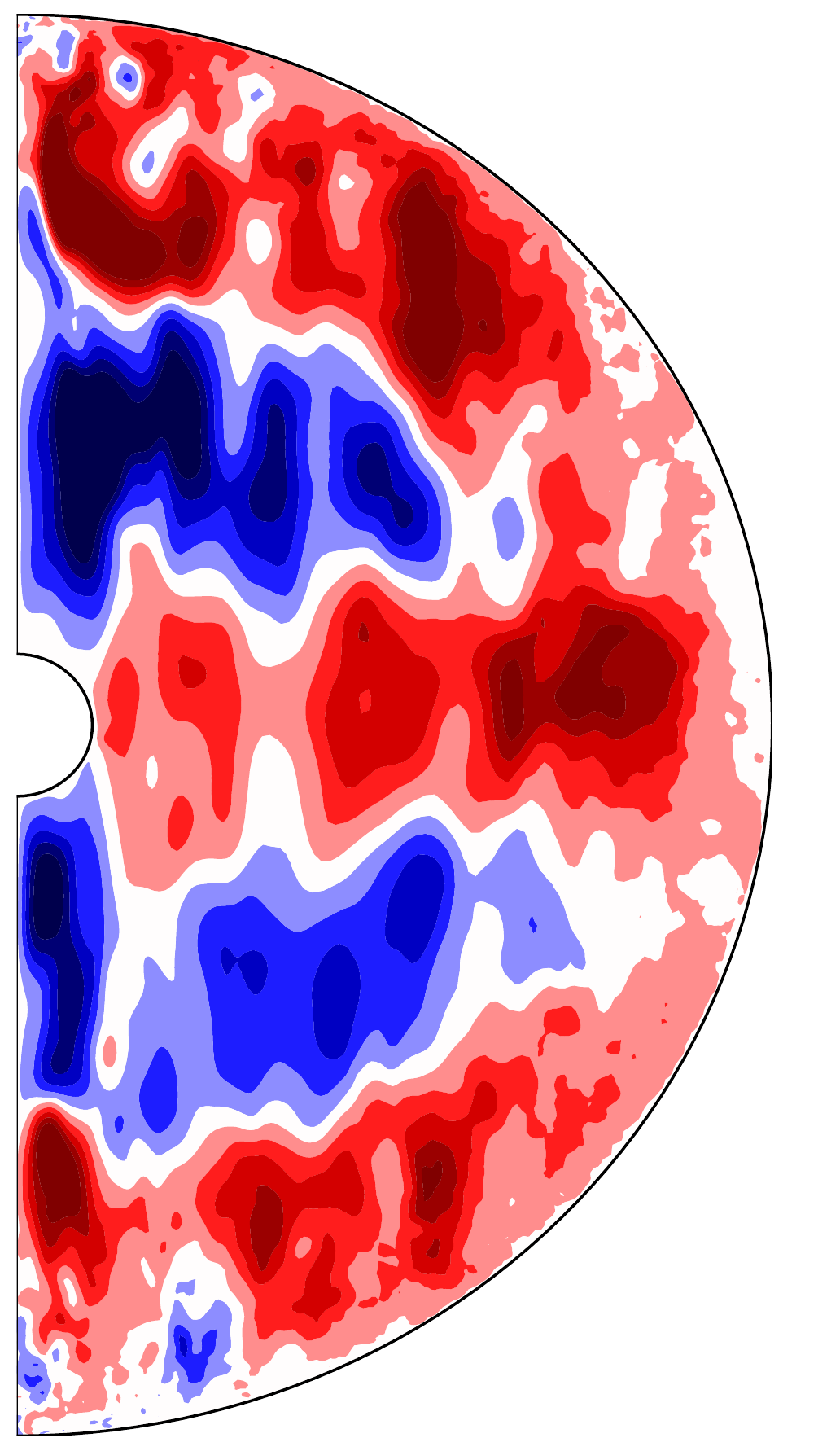} 
\plotone{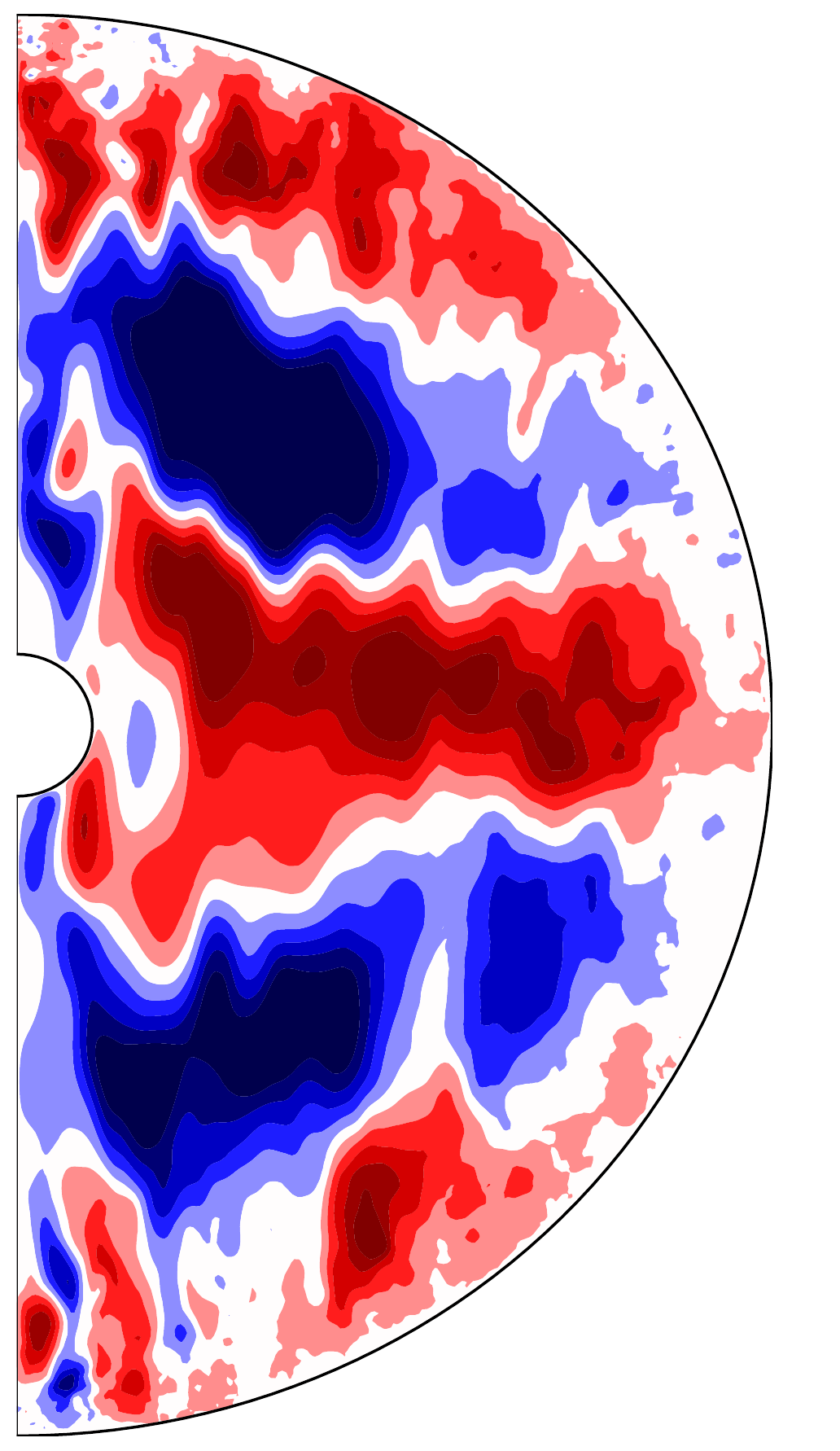} \plotone{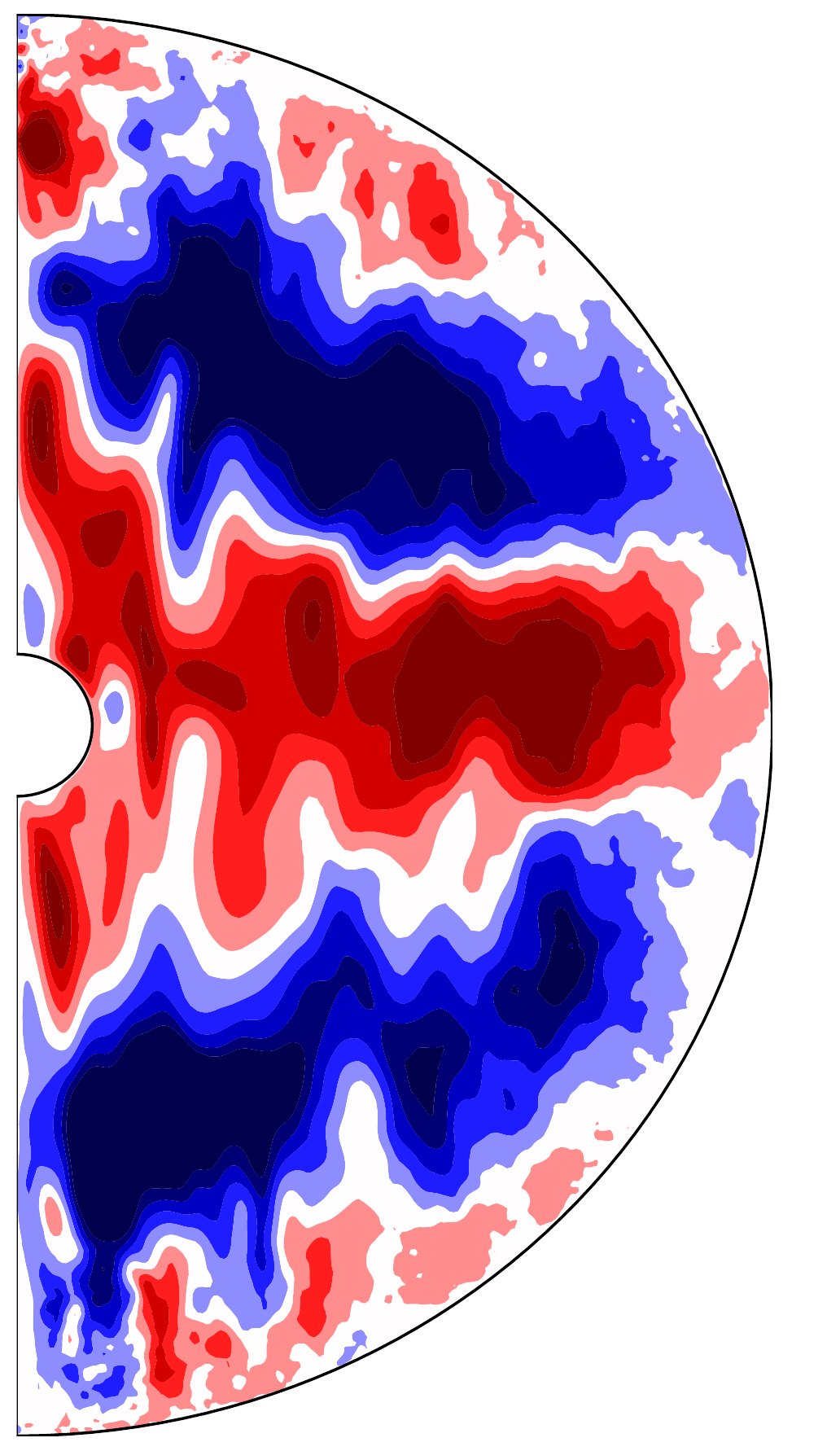} \plotone{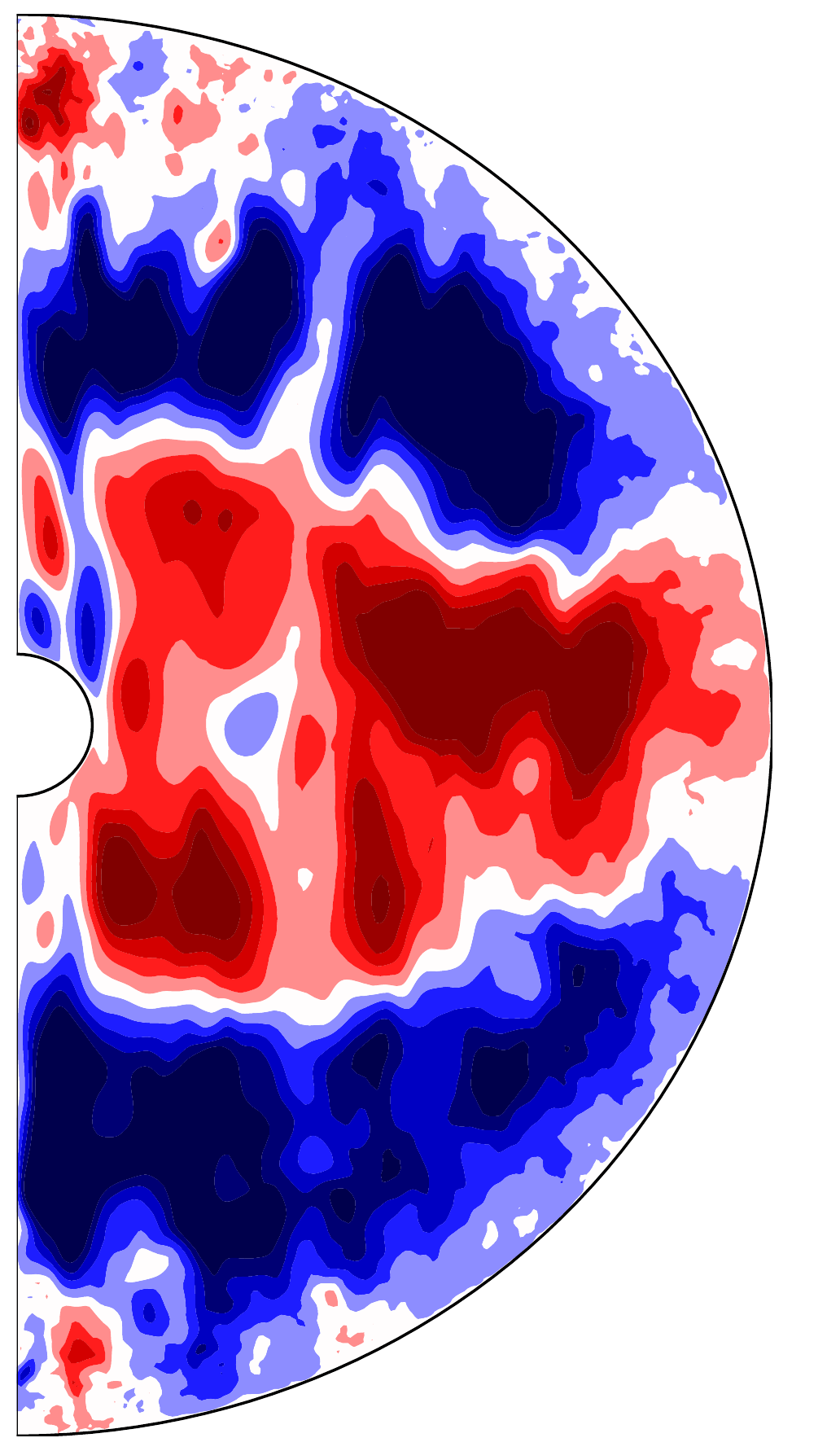} \\
\vspace{0.2cm}
\plotone{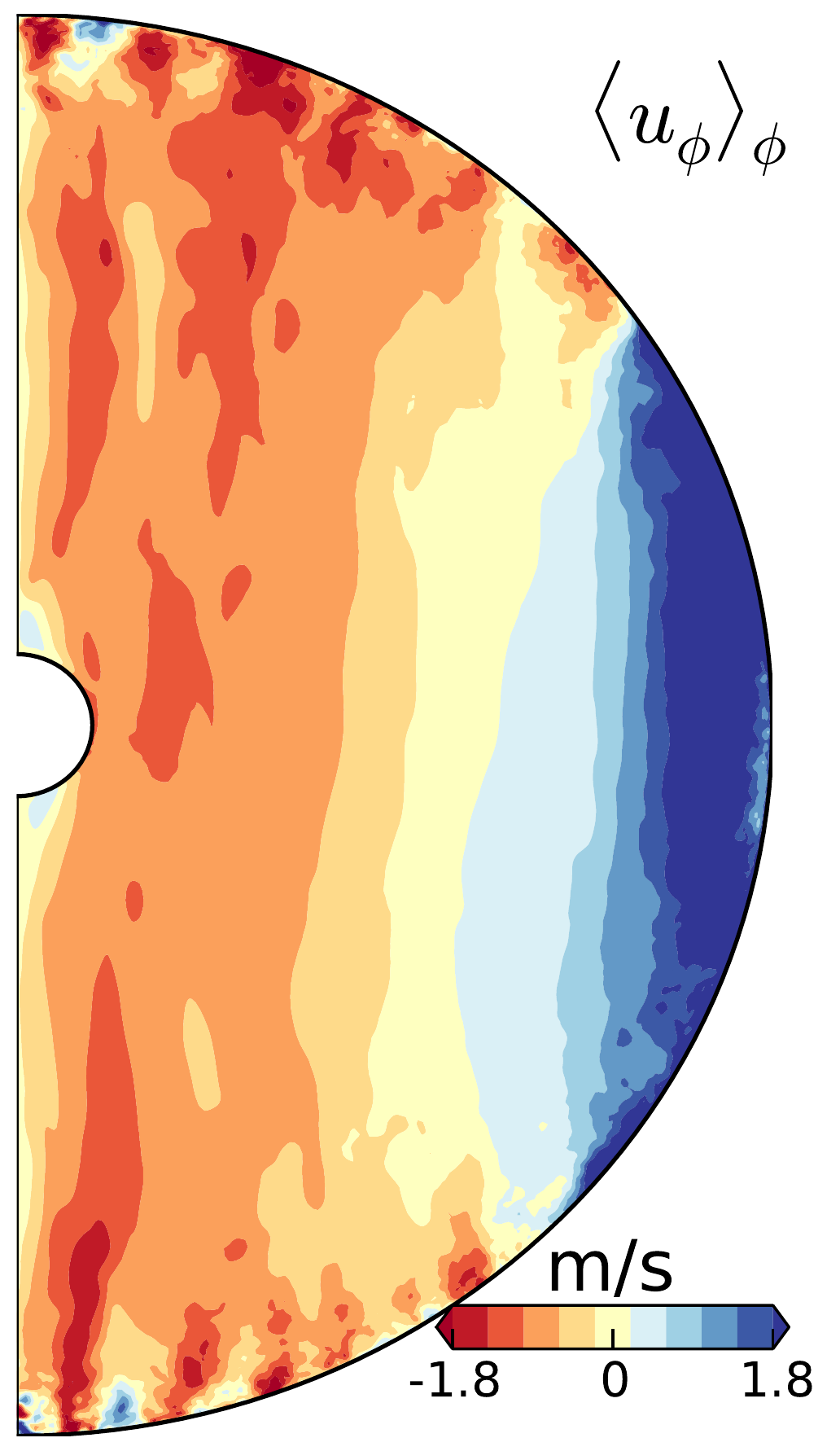} \plotone{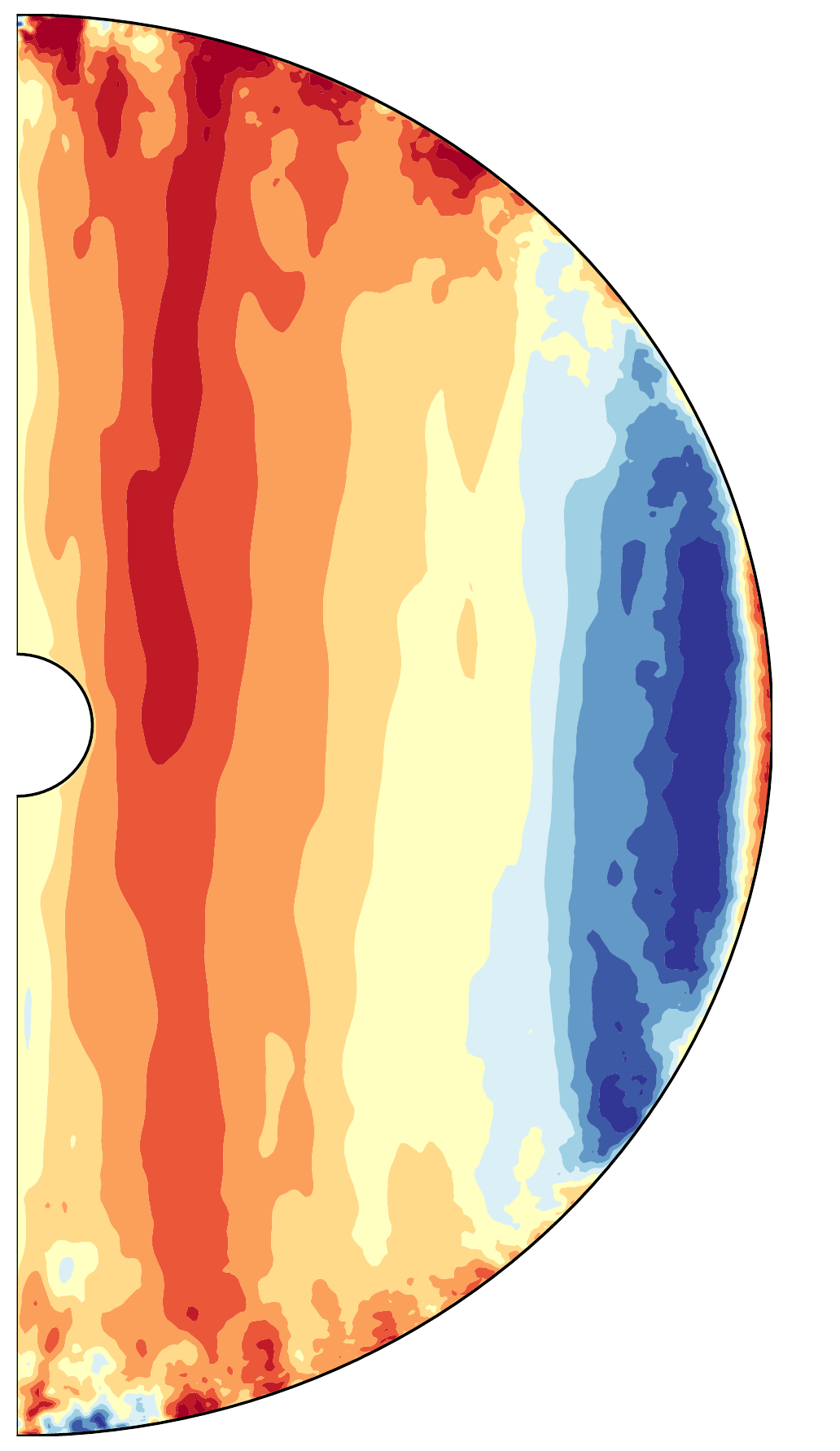} 
\plotone{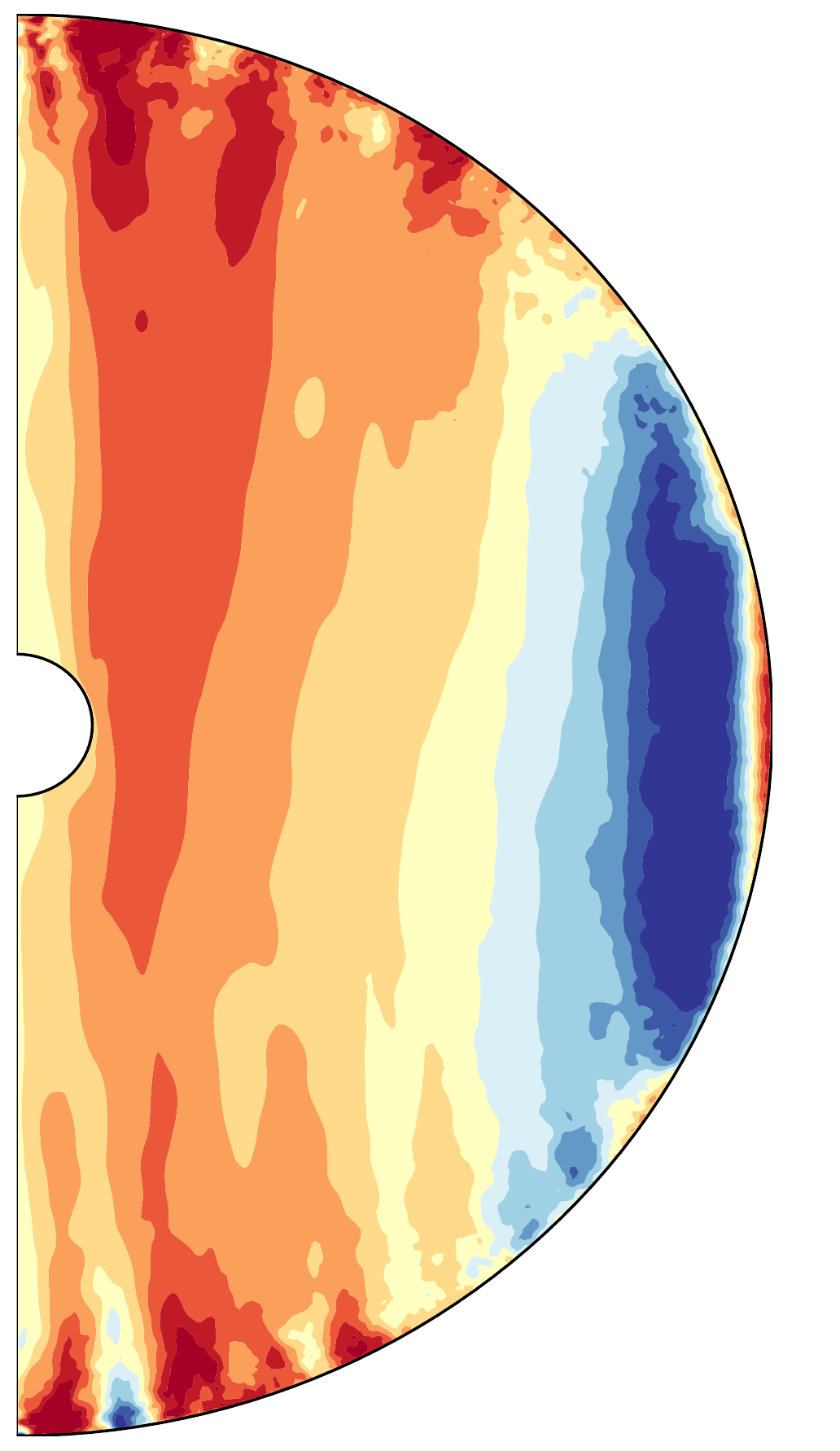} \plotone{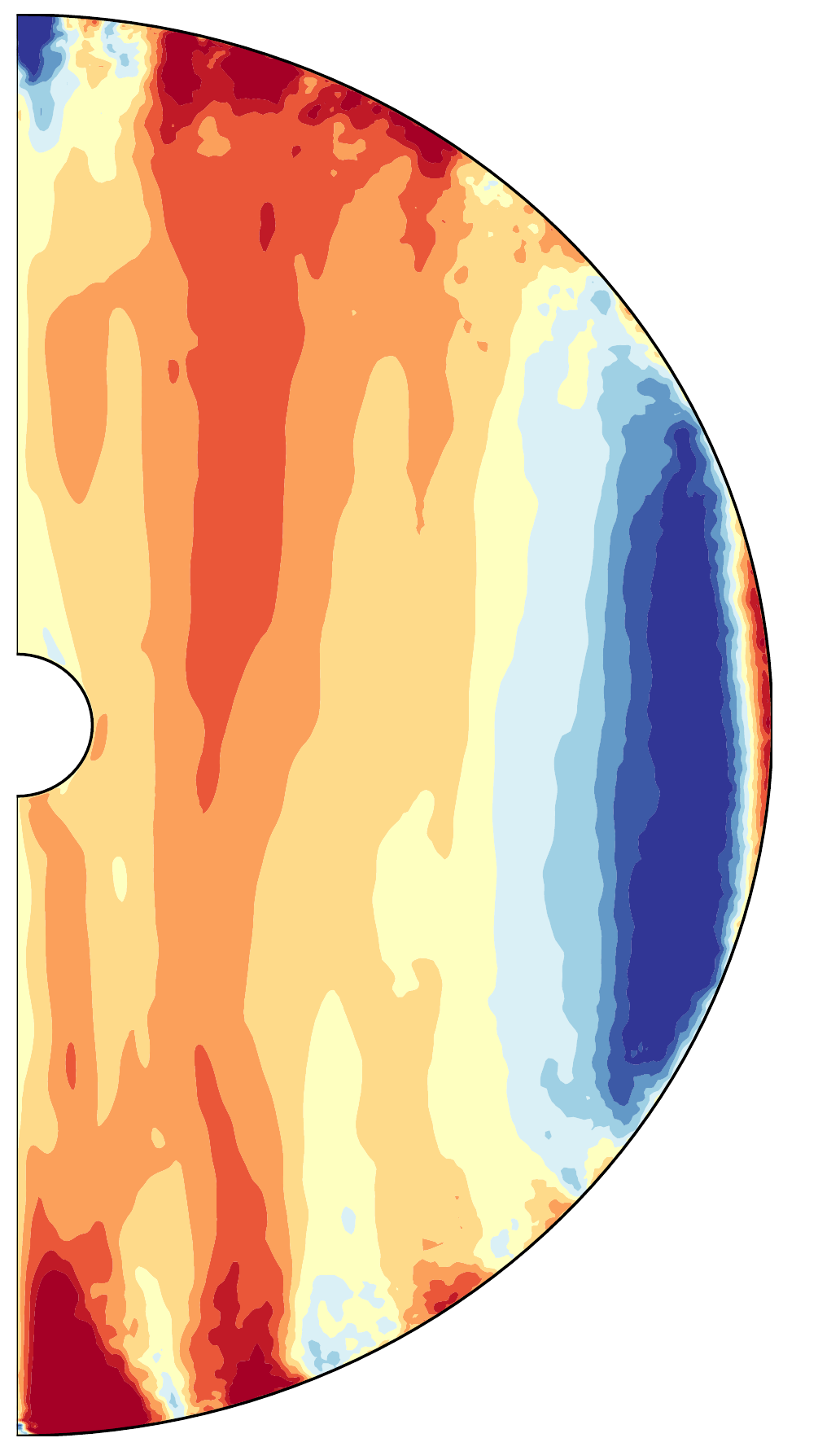} \plotone{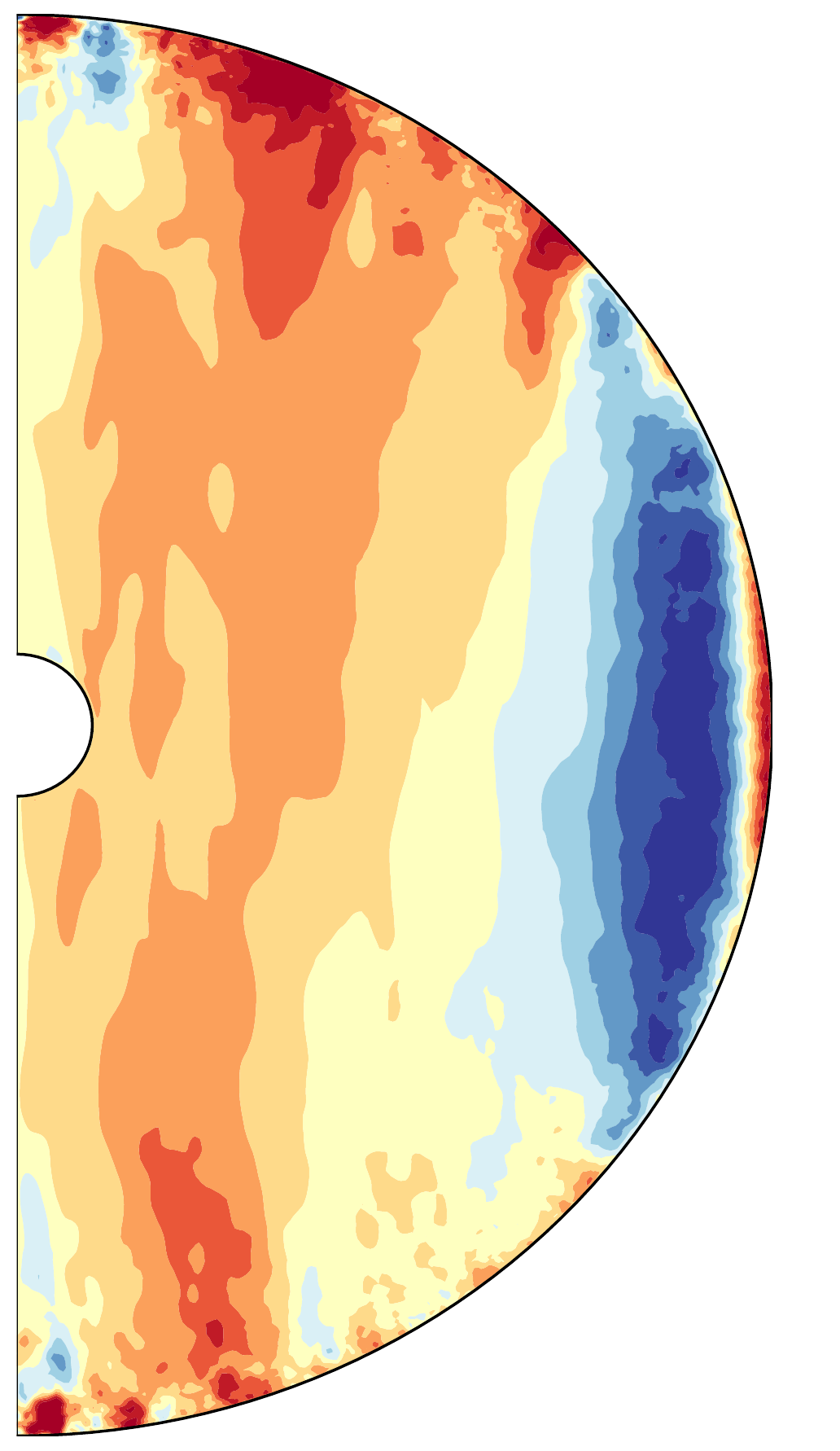}
\caption{Snapshots of longitudinally averaged longitudinal magnetic field and longitudinal velocity as a function of radius and latitude. The snapshots are separated by about 5 rotations. \label{fig2}}
\end{figure*}

Analysing the evolution of the radial and azimuthal magnetic field, presented in panel Figure \ref{fig1}(b) and (c) as Butterfly diagrams, reveals the source of the modulations in the magnetic field time series. The simulation is undergoing magnetic cycles that lead to  repeated reversals of the magnetic polarities at different latitudes as time progresses. The evolution is rather complex: some cycles produce magnetic features that are equatorially symmetric, i.e.~with the same polarities at opposite latitudes in northern and southern hemispheres, while some other cycles produce equatorially anti-symmetric features. There is also one cycle where the northern hemisphere did not produce much axisymmetric field (at around 400 rotations). Note that mean-field dynamo studies, which parametrize the differential rotation and the $\alpha$ effect due to helical convection, also predict magnetic cycles in FC M-stars, including for those that rotate rapidly \citep{shulyak2015, pipin2016}.

In panel (b), let us concentrate on one of the magnetic polarity features (say, around 100 rotations). There are two distinct branches migrating from the polar and the equatorial latitudes to the mid-latitude regions. However, there are also cycles where the polar branch is stable and does not migrate. A similar behaviour is present in the radial magnetic field (panel (c)) but to a lesser extent. The temporal evolution of the axisymmetric components of the longitudinal magnetic field and velocity in a typical cycle with an equatorially symmetric structure is presented in Figure \ref{fig2}. Magnetic field features originate in the deep interior and migrate in radius to replace the magnetic field with opposite polarity in the outer layers. Therefore, the entire convection zone participated in building up a cycle. The differential rotation (DR) also varies substantially through the cycle (lower panel in Figure \ref{fig2}). In our earlier simulation with smaller Rossby number \citep{yadav2015apjl}, DR was quenched in most of the convection zone and persisted only in the outer layers near the equator (see Fig.~1d in \cite{yadav2015apjl}). In this simulation, however, differential rotation exists in the deeper convection zone as well as at higher latitudes. In a thin layer close to the equator of the outer surface DR is antisolar. This feature is likely to grow at higher Rossby numbers and eventually overwhelm the solar-like DR in the equatorial latitudes \citep{gastine2013zonal}. However, in the present model, the thin antisolar feature near the equator is rather intermittent and appears only occasionally. For the most part of the simulation DR is solar-like in the equatorial regions.

Magnetic cycles qualitatively similar to the one we find here have been reported in several earlier studies, although mostly in thinner convection zones than ours \cite[e.g., see][]{gastine2012a, kapyla2013, warnecke2014, jones2014, augustson2015, duarte2016, raynaud2016}. It has been argued that such dynamo solutions are Parker-waves that are driven by the strong DR produced in the convection zone \citep[for example, see][]{busse2006, gastine2012a, warnecke2014}. Indeed such cyclic dynamo solutions are always accompanied by a larger DR \citep{busse2006, browning2008, gastine2012a, yadav2013a} as compared to their quasi-steady dipole-dominant counterparts where DR is highly quenched \citep{aubert2005, gastine2012a, yadav2013a, yadav2015spots}.

\begin{figure*}
\epsscale{0.9}
\plotone{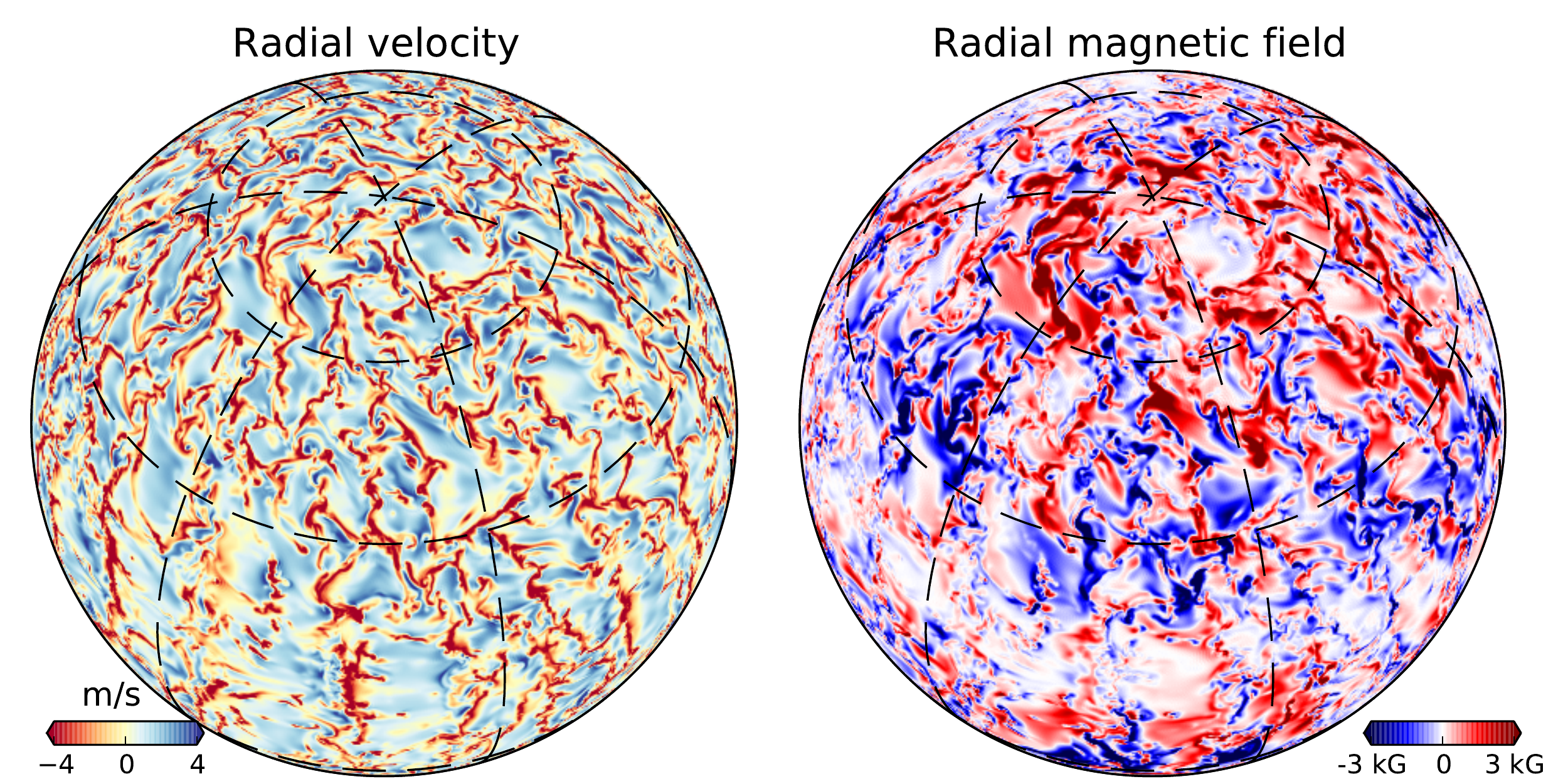}
\caption{The radial velocity and the radial magnetic field on a layer close  to the outer boundary ($r=0.97r_o$) of the simulation. Note that this snapshot is from a simulation segment that was run on a higher resolution grid (2048 in longitude, 1024 in latitude, 161 in radius). \label{fig3}}
\end{figure*}

Along with producing a strong axisymmetric magnetic field, the simulation also shows localized bipolar magnetic structures that reach high field strengths of about $\pm 3$ kG. Such field strengths should be enough to produce starspots. As shown in Figure \ref{fig3}, these regions are distributed almost uniformly across the surface. The bipolar features are mostly formed in narrow convective downwellings where a relatively high level of turbulence is produced due to the converging and colliding flows.


\section{Summary and discussion}
In this Letter we investigate the effect of Rossby number (or rotation period) on the dynamo mechanism in a turbulent simulation of fully convective M-stars. We build on our recent modelling strategy that showed that anelastic dynamo simulations can self-consistently reproduce magnetic field properties observed at rapidly rotating  ($Ro\lesssim 0.1$) fully convective M-stars \citep{yadav2015apjl}. Here, we simulate a model with a rotation period similar to Proxima Centauri (about 83 days). The Rossby number in this slowly rotating setup is about 0.25. The increased Rossby number fundamentally changes the nature of the dynamo solution and generates coherent magnetic cycles. These cycles have a complex temporal evolution where magnetic field structures migrate to mid-latitudes from poles and equator. Some of these cycles produce a magnetic field that is symmetric about the equator  while some others produce antisymmetric fields. Our simulation demonstrates that {\em large Rossby numbers may promote regular activity cycles in fully-convective stars.}

It is instructive to contrast our earlier simulation with a relatively low Rossby number of about 0.05 \citep{yadav2015apjl} and the current setup with a higher value of 0.25. At low Rossby numbers: 1) The magnetic field is non-cyclic and dipole dominated, i.e.~it is mostly concentrated on mid and high latitudes. 2) A substantial portion (about 30\%) of the total magnetic energy resides in the axisymmetric part. 3) The differential rotation is highly quenched in most of the convection zone. Lacking a tachocline and differential rotation, only the helical turbulence is sustaining the dynamo. Such $\alpha^2$ type dynamos might be similar to the dynamo operating in the Earth's convection zone \citep{christensen2009, yadav2013b}. At larger Rossby numbers: 1) The differential rotation persists throughout the convection zone and at higher latitudes. 2) The amount of magnetic energy in the axisymmetric component decreases to about 13\%. 3) The dynamo starts producing magnetic cycles. 4) The large-scale magnetic field is widespread across different latitudes. Due to the  differential rotation in the convection zone ($\Omega$-effect) such dynamos can be categorized as being of the $\alpha\Omega$ type. Here, both large scale shear and helical turbulence participate in sustaining the dynamo. Based on these inferences from our simulations, we may speculate that {\em fully convective M-stars will transition from steady $\alpha^2$-type dynamos to cyclic $\alpha\Omega$-type as the Rossby number increases}.

Using a rotation period of 83 days and a convective time scale of about 150 days \citep{wright2011}, the resulting Rossby number in Proxima Centauri is about 0.5, twice larger than in our simulation. However, it must be kept in mind that the simulation convective time scale is based on the actual velocities generated in the model, whereas the observationally constrained convective time scale is an empirical estimate calculated by minimizing the scatter of data around the rotation-activity relationship \citep{wright2011}. Therefore, a mismatch of a factor of a few between the theoretical and observationally inferred Rossby numbers should not be surprising. Given these uncertainties in the Rossby number, the agreement of the 9-year periodicity found in our model with the observed 7-year activity cycles at Proxima Centauri is remarkable. An intrinsic assumption here is that the activity will peak when the mean magnetic field strength on the simulation surface peaks. Furthermore, as mentioned earlier, we find that a larger Rossby number tends to distribute the large-scale magnetic field across different latitudes. Although most M-stars will probably have numerous small starspots distributed almost uniformly across the surface, one can imagine that if the large-scale magnetic field is strong enough it may lead to relatively larger starspots or starspot groups. If true, then we can expect that {\em large starspots/starspot groups will form at latitudes ranging from the equator to the poles on Proxima Centauri}, unlike the case in rapidly rotating FC M-stars where large starspots/starspot groups preferentially form at high latitudes \citep[e.g.][]{barnes2015}.

To pin down the mechanism that generates the magnetic cycles in our model we need to analyze the current simulation in detail and  perform additional simulations, especially at different Rossby numbers (different Rayleigh numbers and Ekman numbers) and at different magnetic diffusivities (different magnetic Prandtl numbers). Such an exercise will reveal how much the cycle length is affected by these fundamental parameters. The role of the fluid viscosity and the magnetic Reynolds number on the dynamo mechanism is also an outstanding issue which needs to be tackled with future studies. Due to the aforementioned limitations of numerical simulation, the robustness of the cycle period we get in our study can be justifiably questioned. However, since our simulations probe Rossby numbers similar to those found in FC M-stars, our conjecture that low $Ro$ convection produces quasi-steady dipole dominant dynamos and higher $Ro$ leads to cyclic behavior is probably a more robust result.

To better constrain the dynamo theory for fully convective stars like  Proxima Centauri, the observational landscape needs to be improved substantially in the coming years. Long term monitoring campaigns are now beginning to find near-by slowly rotating fully convective M-stars with rotation periods in the 70 to 100 days range \citep{irwin2011, newton2016}. An essential, albeit labor intensive, exercise is to monitor such M-stars on the time scale of decades, similar to the one carried out for Sun-like stars \citep{baliunas1995}. Due to slower rotation, the Zeeman-Doppler Imaging technique \citep{donati2009} will not be very helpful in revealing the large-scale morphological features of the magnetic field on Proxima Centauri. Therefore, detailed theoretical models that match the available observations will be essential for characterizing the Proxima Centauri system.

\acknowledgments
We thank the referee for providing several constructive comments that improved the manuscript. RKY is supported by NASA {\em Chandra} grant GO4-15011X and SJW is supported by NASA contract NAS8-03060. Simulations were performed at RZG and GWDG. This work also used the Extreme Science and Engineering Discovery Environment (XSEDE), which is supported by National Science Foundation grant number ACI-1053575 \citep{xsede}.

\bibliographystyle{apj}

\end{document}